\documentstyle[12pt]{article}
\input epsf
\newcommand{\NPB}[3]{\emph{ Nucl.~Phys.} \textbf{B#1} (19#2) #3}   
\newcommand{\PLB}[3]{\emph{ Phys.~Lett.} \textbf{B#1} (19#2) #3}   
\newcommand{\PRD}[3]{\emph{ Phys.~Rev.} \textbf{D#1} (19#2) #3}   
\newcommand{\PRL}[3]{\emph{ Phys.~Rev.~Lett.} \textbf{#1} (19#2) #3}

\def\dalemb#1#2{{\vbox{\hrule height .#2pt
        \hbox{\vrule width.#2pt height#1pt \kern#1pt
                \vrule width.#2pt}
        \hrule height.#2pt}}}

 \def\bd{\begin{document}} \def\ed{\end{document}}
\def\ds{\documentstyle} \let\fr=\frac \let\bl=\bigl \let\br=\bigr
\let\Br=\Bigr \let\Bl=\Bigl 
\let\bm=\bibitem
\let\na=\nabla
\let\pa=\partial \let\ov=\overline
\def\ie{{\it i.e.\ }} 
\newcommand{\pr}{\paragraph{}}
\def\beq{\begin{equation}}
\def\eeq{\end{equation}}
\def\beqa{\begin{eqnarray}}
\def\eeqa{\end{eqnarray}}

\newcommand{\ba}{\begin{array}}
\newcommand{\ea}{\end{array}}
\newcommand{\td}{\tilde}
\newcommand{\norsl}{\normalsize\sl}
\newcommand{\ns}{\normalsize}
\newcommand{\refs}[1]{(\ref{#1})}
\def\simlt{\mathrel{\lower2.5pt\vbox{\lineskip=0pt\baselineskip=0pt
           \hbox{$<$}\hbox{$\sim$}}}}
\def\simgt{\mathrel{\lower2.5pt\vbox{\lineskip=0pt\baselineskip=0pt
           \hbox{$>$}\hbox{$\sim$}}}}

\def    \ccii          {C_i^{5_i}}
\def    \ccci          {C_i^{5_3}}
\def    \ccia          {C_1^{5_i}}  
\def    \ccib          {C_2^{5_i}}  
\def    \ccic          {C_3^{5_i}}  
\def    \ccij          {C_{j}^{5_{i}}}
\def    \ccik          {C_{k}^{5_{i}}}
\def    \ccjk          {C_{j}^{5_{k}}}
\def    \cckj          {C_{k}^{5_{j}}}
\def    \cca           {C_1^{5_1}}
\def    \ccb           {C_2^{5_2}}   
\def    \cccc          {C_3^{5_3}}
\def    \ccbc          {C_2^{5_3}}
\def    \cccb          {C_3^{5_2}}
\def    \ccac          {C_1^{5_3}}
\def    \ccab          {C_1^{5_2}}
\def    \ccba          {C_2^{5_1}}
\def    \ccca          {C_3^{5_1}}
\def    \cni           {C_i^{9}}  
\def    \cna           {C_1^{9}}
\def    \cnb           {C_2^{9}} 
\def    \cnc           {C_3^{9}}
\def    \cnci          {C^{9 5_i}}
\def    \cncj          {C^{9 5_{j}}}   
\def    \cnck          {C^{9 5_{k}}}
\def    \cnca          {C^{9 5_1}}  
\def    \cncb          {C^{9 5_2}} 
\def    \cncc          {C^{9 5_3}}
\def    \ccjck         {C^{5_{j} 5_{k}}}
\def    \ccick         {C^{5_{i} 5_{k}}}
\def    \ccicj         {C^{5_{i} 5_{j}}}
\def    \ccacb         {C^{5_1 5_2}}
\def    \cccca         {C^{5_3 5_1}}
\def    \ccbcc         {C^{5_2 5_3}}
\begin{document}
\thispagestyle{empty}
\rightline{CERN--TH/99-359}
\rightline{\large\sf hep-ph/9911517}
\rightline{\large November 1999}
\vskip 1.0truecm
\centerline{\Large\bf A Note on New Sources of Gaugino Masses }
\vskip 1.truecm
\centerline{ {\large\bf Karim
Benakli}\footnote{e-mail:Karim.Benakli@cern.ch}}
\vskip .5truecm
\centerline{{\it CERN Theory Division
 CH-1211, Geneva 23, Switzerland}}
\vskip .5truecm

\vskip 1.truecm
\centerline{\bf\small ABSTRACT}
\vskip .5truecm

In IIB orientifold models, the singlet twisted moduli appear in the 
tree-level gauge kinetic function. They might be responsible for
generating gaugino
 masses if they acquire non--vanishing $F$-terms. We discuss some aspects
of this 
new possibility, such  as the size of gaugino masses and their
non-universalities.
A possible brane setting 
is presented to illustrate the usefulness of these new sources.

\hfill\break
\vfill\eject

Supersymmetry breaking  is a major issue in superstring and M-theory. It
is
for instance necessary  to lift the  degeneracy of vacua. For 
phenomenological applications,
supersymmetry breaking will provide mass splitting between supersymmetric 
partners, explaining why these have not been observed in nature yet. 
 The precise dynamics  involved in the generation of 
such masses
is still unknown, but one can use  a phenomenological parametrization
which turns out to be useful for many purposes dealing with low-energy
predictions.

 For weakly coupled heterotic strings, such a line of ideas was
advocated  in \cite{bim}, where  non--vanishing
$F$-terms were assumed for the moduli fields $S$ (dilaton) and
$T_i$ (associated to the Kahler  structure of the compact internal
space). The gauge groups originating from  reduction of the
ten-dimensional gauge symmetry have a universal  tree-level coupling.
 Non-universalities of couplings and gaugino masses arise
at one-loop through a $T_i$ dependence of threshold corrections. 
 
Another convenient framework to pursue these investigations  is provided
by 
type IIB orientifolds. Soft terms for such
compactifications have been discussed  in \cite{IMR}. It was
noticed that non-universal gaugino masses could be generated if for
instance different parts of the standard model gauge group originated
from different sets of branes \cite{IMR,Kane}. To allow unification of 
gauge 
couplings one would then need to
construct models where the moduli, controlling the gauge couplings on 
different branes, 
get potentials with minima at the same value. 
Here we will address another origin for soft terms: twisted moduli related 
to blowing-up modes.

The IIB orientifolds are obtained as 
compactifications on three tori $T^1, T^2, T^3$ on which different points 
are identified under a discrete symmetry $Z_N$, which leads to a set of 
fixed points. Requiring 
 ${\cal N}=1$ supersymmetry and Poincar\'e invariance in four dimensions 
allows  the presence  
of 9- and 5-branes (equivalently under $T$-dualities 3- and 7-branes).

The space group action of the orbifold is defined by some twist
eigenvector
$v=(v_1,v_2,v_3)$. In the sector twisted  by $\theta^k$ the orbifold group
acts as $X_i \rightarrow \theta^k X_i$,  $\theta = \exp{2\pi i v\cdot J}$,
where $J=(J_1,J_2,J_3)$, 
with $X_i$ and $J_i$ the coordinate and generator of rotation in the
$i$-th torus respectively.
For a given twist $\theta^k$ one finds $\prod_{i=1}^3 4 \sin^2 \pi kv_i$
fixed points that we label by an index $f$. In a similar way, 
the orbifold acts also on the Chan-Paton factors through some twist
parametrized by a vector $V_a$ with model-dependent fractional entries
$l/N$. 
In the case of even $N$,  some sets of $D5_i$-branes are present, sitting
 at the origin  $X_j =X_k =0$ in the 
$j \neq i$ and $k \neq i$ complex planes. We label by an index $p_i$ the
 $4\sin^2 \pi k v_i$ fixed
points located in the world-volume of the 
$D5_i$-branes.

In addition  to the dilaton $S$ and to the  three moduli $T_i$, $i=1,2,3$,
parametrizing the Kahler structure (volume) of the tori, there are the
twisted moduli $Y^k_f$ associated 
to blowing-up the orbifold singularities $f$ due to a twist $\theta^k$
\footnote{ We have changed notation from the usual $M^k_f$ to avoid
confusion with masses.}. The new feature in IIB orientifolds is that
these moduli couple at tree-level to gauge kinetic terms. The gauge
kinetic functions for gauge fields on the D9- and D5-branes are
given by \cite{S,IIB,ABD}

\beqa
f^9_{b} & = & S + \frac{1}{N}\sum_{k=1}^{[N-1/2]} 
\frac {\cos 2\pi k V^9_{b}}{\prod_{i=1}^3 2 \sin \pi k v_i}
\; \sum_{f} Y^k_{f}  \nonumber \\
f^5_{ia} & = & T_i + \frac{1}{N}\sum_{k=1}^{[N-1/2]} 
\frac {cos 2\pi k V^5_{ia}}{2 \sin \pi k v_i} \; \sum_{p_i} Y^k_{p_i}.
\label{funcfs}
\eeqa

In general (\ref{funcfs}) results in different independent linear 
combinations of $Y_{ai}$ of the $Y^k_{i}$ for each of the    
gauge kinetic function corresponding to gauge groups $G_a$. 
So $F$-terms for the twisted  moduli 
$Y^k_{i}$ will be a new source of tree-level gaugino masses:

\beq
M_{a} =  \sum_{i} c_a^i M_{Y_{ai}},
\label{gau1}
\eeq
where $M_{Y_{ai}}$ are the contributions of different $Y_{ai}$
and the coefficients $c_a^i$ are model-dependent.
We see that the gaugino masses and the associated complex 
phases could be  non-universal in these models.

The cases of odd $N$ lead to a drastic simplification.
Only one  linear combination, which we denote as $Y$, of the twisted
moduli  appears in 
the gauge kinetic function.
The coefficient of the dependence for the group $G_a$ is given by 
the beta-functions $b_a$ of the running of the corresponding gauge 
coupling \cite{mirage}:

\beq
f^9_{a}  =  S + \frac {b_a}{2} Y.
\label{funcf9}
\eeq

In the absence of an $F$-term for $S$ but for $Y$,  a tree--level 
gaugino mass  proportional to the one-loop beta-function coefficient 
will be generated (using the convention of \cite{bim}):

\beq
M_{a}  = \frac {\sqrt{3}}{2}\; \frac { b_a g_a^2}{16 \pi^2} \; m_{3/2} \;
e^{-\alpha_Y}
\; (K_Y^Y)^{-1/2} = \sqrt{\frac {3}{8}}\; \frac { b_a g_a^2}{16 \pi^2}\;
m_{3/2} \; e^{-\alpha_Y}
\label{gau2}
\eeq
where we have used ${\rm Re} f^9_{a} = {8 \pi^2/g_a^2 }$ with $g_a$ the 
four-dimensional gauge coupling. In (\ref{gau2}), $\alpha_Y$ is the
complex phase, 
$K$ is the Kahler potential which we assumed in the second equality to be
given by $(Y+\bar Y+\cdots)^2$. The fact that 
the form of the gaugino masses is similar to a one-loop form can be traced 
back to 
the fact that the dependence on $Y$ is there to cure sigma-model  
anomalies \cite{kah}. One-loop contributions to gaugino masses could be
important 
in this case\footnote {A nice discussion of such effects might be found
for example in \cite{poppitz}.}.

The relation (\ref{gau2}) means that the gauginos have non-universal
masses 
but 
a unique 
phase. The beta-functions coefficients $b_a$ take into account all 
the states 
that are massless at the string scale. If these are identified with the
low-energy 
ones, one then has the low-energy prediction:

\beq
\frac {M_{3}} {b_3} = \frac {M_{2}} {b_2} = \frac {M_{1}} {b_1},
\label{gau3}
\eeq
where $M_3$, $M_2$ and $M_1$ are the gaugino masses associated with the
 $SU(3)$, $SU(2)$ and $U(1)$ factors of the standard model.

The presence of both  $F_S$ and $F_Y$ will obviously  lead to
non-universal gaugino masses with two independent phases, one of which
could be chosen to vanish. The $F_S$ is expected to dominate because of
the 
coupling constant suppression of the $F_Y$.

Does this non-universality also mean
that  gauge unification is lost? The crucial issue here is that
although we have  used non-vanishing $F_{Y^k_{i}}$, we have made
no assumption on the  vacuum expectation values of $Y^k_{i}$ moduli
themselves. In fact, to be more precise, the gauge kinetic function is
given in the string basis by linear multiplets $l$ and $y_{as}$:

\beqa
f_{a}\ &  = & \ \frac{1}{l} + \sum_{s} c_{as} y_{as},
\label{kineticf}
\eeqa
where $c_{as}$ are model-dependent constants. Under linear-chiral duality, 
$l$ is associated with the dilaton while $y_{as}$ are associated with the 
$Y^k_{i}$ moduli. It was argued in
\cite{ABD} that  the latter modulus $y_{as}$  should have a
vanishing\footnote{Supersymmetry breaking could lead to  vevs
$y_{as}$, but these should remain very small to keep the orientifold
picture valid.} vev to be in
the orientifold limit, where our results are valid. This ensures
automatic unification.

Let us turn to some brane setting to illustrate how this new possibility 
can be useful.

In general one might have 9-branes and  three types of $5_i$--branes
corresponding to the  different choices $T^i$ of the torus on which
the $5$-branes are wrapping. There are three kind of charged  states
that originate from open strings stretched between 9-branes denoted as
(99) states, those stretched between 5-branes denoted as ($5_i5_j$)
and those stretched between 5-branes and 9-branes denoted as
($5_i9$).  The (99), $(5_i5_i)\equiv (55)_i$  strings give rise to
gauge vector multiplets of the corresponding gauge group $G^9$ and
$G^5_i$, respectively. They also lead to chiral multiplets ($99$) and
$(55)_i$ charged only under $G^9$ and $G^5_i$ respectively.  In
contrast,  the ($5_i5_j$) lead to chiral fields  charged
under  both  $G^5_i$ and $G^5_j$, while ($5_i9$) open strings lead to
 chiral superfields charged
under  both  $G^5_i$ and $G^9$.

Suppose that the standard model gauge symmetry originates from
9-branes.  We  also assume that there are two (or three, but the the last
one
 plays  no role) sets of D5-branes: $5_1$ located at $X_2 =X_3=0$
and $5_2$ located at $X_1 =X_3=0$. The gauge coupling on the two sets are
given by:

\beqa
f^5_{1a} & = & T_1 + \frac{1}{N}\sum_{k=1}^{[N-1/2]} 
\frac {cos 2\pi k V^5_{1a}}{2 \sin \pi k v_1} \; \sum_{p1} Y^k_{p1}
 \nonumber \\
f^5_{2a} & = & T_2 + \frac{1}{N}\sum_{k=1}^{[N-1/2]} 
\frac {cos 2\pi k V^5_{2a}}{2 \sin \pi k v_2} \; \sum_{p2} Y^k_{p2}
\label{f512}
\eeqa

Consider the $5_1$-brane to be a hidden sector where non-perturbative
effects break supersymmetry breaking and generate $F$--terms for some
of  the $Y^k_{p1}$ moduli. This could arise from gaugino condensation (or
string-scale breaking as the brane--antibrane models of \cite{bbar} if the
string
scale is at an intermediate region \cite{int}), which leads to a potential
which goes as $e^{- c/g^2_{1a}}$, which depends on the  $Y^k_{p1}$ and
could lead to $F$-terms for the latter. The  $9$-brane standard
model gauge  kinetic function involves all the twisted moduli and 
will thus have the corresponding gaugino masses generated at tree-level.

We identify the standard model matter fields as coming from
($5_29$) open strings. These feel only the $Y^k_{p2}$ moduli, which
share  with the $Y^k_{p1}$ set only the modulus $Y^k_{0}$ associated
to the blowing-up  mode of the origin $X_1 = X_2 = X_3=0$. If
$F_{Y^k_{p1}} \neq 0$ for  $p_1 \neq 0$ and $F_{Y^k_{0}} = 0$, then the
scalar soft masses will be generated at one-loop only, mediated by
gaugino masses. This might provide a brane realization for the
scenario\footnote{The nice phenomenological peculiarities of soft terms as
suggested in \cite{ ann} were also present in \cite{pomarol}. The
low-energy predictions are also similar to \cite{noscale}. I thank A.
Pomarol for stressing these points  to me.} proposed
in \cite{ ann}. However, here the gaugino masses are generically
non-universal. A $\mu$-term of the same order as the gaugino masses will
be generated through a Kahler potential \cite{GM} if there is a coupling
$Y^k_{p2} H_1 H_2$. Such a term is expected for instance for
the case of compactifications  of  the form $(K3 \times T^2)/\Gamma$ 
with a singular $K3$ and $\Gamma$ a discrete symmetry as $Z_N$. 
 Before  compactification on $T^2$ and acting 
with $\Gamma$, it is known from \cite{S} that there are  couplings 
$Y^k_q F^2$ with $F^2$
the six-dimensional gauge field strength from the (99) sector and 
$Y^k_q$ are the twisted
moduli associated with blowing up the $K3$ singularities. Now upon 
the reduction to  some of the gauge-field components will
lead to chiral fields in four dimensions that could be identified with 
the Higgs doublets. It is interesting to look for  explicit string models 
with such properties.

In conclusion, we have seen that in addition to contributions from 
the dilaton $S$ 
and the moduli fields $T_i$, there might be new contributions from twisted
moduli
$Y^k_f$ corresponding to blowing-up modes for the singularities of
IIB-orientifolds.
These are generically present and allow an extension of new possiblities
for 
soft-terms as generic non-universalities of masses and
 phases, as well as the possibility to naturally restrict the tree-level 
soft-terms 
to part of 
the spectrum while generating other masses 
at higher orders.

\vskip1cm
I am grateful to C. Kounnas, Y. Oz,  A. Pomarol and A. Uranga  for useful
discussions. I wish also to thank G. Giudice and M. Quir\'os for comments
on the manuscript.


\begin{thebibliography}{99}

%
\bibitem{bim} A. Brignole,  L.E.~Ib\'a\~nez and C.~Mu\~noz,
hep-ph/9812397;
see also: V. Kaplunovsky and J. Louis, \NPB{422}{94}{57}; J.P.
Derendinger, 
S. Ferrara, C. Kounnas
and  F. Zwirner, \NPB {372}{92}{145}.
%

\bibitem{IMR} L.E.~Ib\'a\~nez, C.~Mu\~noz and  S.~Rigolin, hep-ph/9812397.
%
\bibitem{Kane} M. Brhlik, L. Everett, G.L. Kane and J. Lykken, 
 \PRL {83} {99} {2124};  hep-ph/9908326; 
E. Accomando, R. Arnowitt and B. Dutta, hep-ph/9909333;
T. Ibrahim and P. Nath, hep-ph/9910553.
%
\bibitem{S} A. Sagnotti, \PLB{294}{92}{196}; 
M.R. Douglas and G. Moore, hep-th/9603167.
%
%
\bibitem{IIB} G. Aldazabal, A. Font,  L.E.~Ib\'a\~nez and G. Violero, 
hep-th/9804026;  L.E.~Ib\'a\~nez, R. Rabad\'an and A.M. Uranga,
hep-th/9808139.
%

\bibitem{ABD} I.~Antoniadis,  C.~Bachas and E.~Dudas, hep-th/9906039.
%
%
\bibitem{mirage} L.E.~Ib\'a\~nez, hep-ph/9905349.

%
\bibitem{kah} L.E.~Ib\'a\~nez, R. Rabadan and A.M. Uranga, hep-th/9905098.

%
\bibitem{poppitz} J.A. Bagger, T. Moroi and E. Poppitz, hep-th/9911029. 
%
\bibitem{bbar} I. Antoniadis, E. Dudas and  A. Sagnotti,
hep-th/9908023; G. Aldazabal and A. M. Uranga, hep-th/9908072;
G. Aldazabal, L.E. Iba\~nez and  F. Quevedo, hep-th/9909172.
%
%
\bibitem{int} K. Benakli, hep-ph/9809582;
C. Burgess, L.E. Ib{\'a}{\~n}ez and F. Quevedo,  hep-ph/9810535.
%
%
%
\bibitem{ann} D.E. Kaplan, G.D. Kribs and M. Schmaltz, hep-ph/9911293;
Z. Chacko, M.A. Luty, A.E. Nelson and E. Ponto, hep-ph/9911323.

%
\bibitem{pomarol}   I.~Antoniadis, \PLB{246}{90}{377}; I.~Antoniadis,
C.~Mu\~noz and M.~Quir\'os, \NPB{397}{93}{515}; I.~Antoniadis and
K.~Benakli, 
\PLB{326}{94}{69}; K.~Benakli, \PLB{386}{96}{106}; I.~Antoniadis and
M.~Quir\'os, \PLB{392}{97}{61}; A.~Pomarol and M.~Quir\'os,
\PLB{438}{98}{255};
I.~Antoniadis, S.~Dimopoulos, A.~Pomarol and M.~Quir\'os,
\NPB{544}{99}{503};
A.~Delgado, A.~Pomarol and M.~Quir\'os, \PRD{60}{99}{095008};  I.
Antoniadis, 
G. D'Appollonio, E. Dudas and A. Sagnotti, \NPB {553 }{99}{133}. 
%

\bibitem{noscale} J. Ellis, C. Kounnas and D.V. Nanopoulos, 
\NPB {241}{84}{406}.    

\bibitem{GM} G.F. Giudice and A. Masiero, \PLB {206}{98}{480}. 



\end{thebibliography}
\end{document}